\documentclass[twocolumn,showpacs,amsmath,amssymb,prd,floats,floatfix]{revtex4}
\usepackage{epsfig}

\begin{document}
\title{Deconfining Phase Transition on Lattices with Boundaries at 
       Low Temperature}

\author{Alexei Bazavov$^{\rm \,a,b}$ and Bernd A. Berg$^{\rm \,a,b}$}

\affiliation{ 
$^{\rm \,a)}$ Department of Physics, Florida State University,
  Tallahassee, FL 32306-4350\\
$^{\rm \,b)}$ School of Computational Science, Florida State 
  University, Tallahassee, FL 32306-4120\\
} 

\date{Jan 7, 2007; revised March 19, 2007}

\begin{abstract}
In lattice gauge theory (LGT) equilibrium simulations of QCD are usually 
performed with periodic boundary conditions (BCs). In contrast to that 
deconfined regions created in heavy ion collisions are bordered by the
confined phase. Here we discuss BCs in LGT, which model a cold exterior
of the lattice volume. Subsequently we perform Monte Carlo (MC) 
simulations of pure SU(3) LGT with a thus inspired simple change of 
BCs using volumes of a size comparable to those typically encountered 
in the BNL relativistic heavy ion collider (RHIC) experiment. 
Corrections to the usual LGT results survive in the finite volume 
continuum limit and we estimate them as function of the volume size. 
In magnitude they are found comparable to those of including quarks. 
As observables we use a pseudocritical temperature, which rises 
opposite to the effect of quarks, and the width of the transition, 
which broadens similar to the effect of quarks.
\end{abstract}

\maketitle

\section{Introduction}

At a sufficiently high temperature QCD is known to undergo a phase 
transition from our everyday phase, where quarks and gluons are
confined, to a deconfined quark-gluon plasma. Since the early days 
of lattice gauge theory simulations of this transition have been a 
subject of the field \cite{Tc-old}, see \cite{Tc-now} for reviews. 
Naturally, such simulations focused on boundary conditions (BCs), which 
are favorable for reaching the infinite volume quantum \cite{quantum} 
continuum limit quickly. On lattices of size $N_{\tau}\,N_s^3$ these 
are periodic BCs in the spatial volume $V = (a\,N_s)^3$, where $a$ 
is the lattice spacing. For a textbook, see, e.g., Ref.~\cite{Rothe}.

The physical temperature of the system on a $N_{\tau}\,N_s^3$ 
lattice, $N_{\tau}<N_s$, is given by
\begin{equation} \label{LGT_T}
  T = \frac{1}{a\,N_{\tau}} = \frac{1}{L_{\tau}}
\end{equation}
where $a$ is the lattice spacing. In this paper we set the physical 
scale by \cite{note},
\begin{equation} \label{Tc}
  T^c=174\ {\rm MeV}
\end{equation}
for the deconfinement temperature, which is approximately the average 
from QCD estimates with two light flavor quarks \cite{Tc-now} in the 
infinite volume extrapolation.  The relation (\ref{LGT_T}) implies for 
the temporal extension of the system
\begin{equation} \label{Ltau}
  L_{\tau} = a\,N_{\tau} = 1.13\ {\rm fermi}\ . 
\end{equation}
For the deconfinement phase created in a RHIC the infinite volume 
limit $N_s/N_{\tau}\to\infty$ for fixed $N_{\tau}$ and subsequently 
$N_{\tau}\to\infty$ ($L_{\tau}=a\,N_{\tau}$ finite) does not apply. 
Instead we have to take the continuum limit as
\begin{equation} \label{fVclim}
  N_s/N_{\tau} = {\rm finite}\,,~~N_{\tau}\to\infty\,,~~L_{\tau}~~
  {\rm finite}\,,
\end{equation}
and periodic BCs are incorrect because the outside is in the confined 
phase at low temperature. Details are discussed in the next section. 

In collisions at the BNL RHIC \cite{RHIC} one expects to create an 
ensemble of differently shaped and sized volumes, which contain 
the deconfined quark-gluon plasma. The largest volumes are those 
encountered in central collisions. A rough estimate of their size 
is
\begin{eqnarray} \nonumber
  & \pi\times (0.6\times {\rm Au\ radius})^2\times c\times 
  ({\rm expansion\ time}) & \\ \label{BNLV}
  & = (55\ {\rm fermi}^2)\times ({\rm a\ few\ fermi}) &
\end{eqnarray}
where $c$ is the speed of light. To imitate this geometry, one may 
want to model spatial volumes of cylindrical and other geometries.
In our exploratory study we do not try to imitate a realistic ensemble 
of deconfined volumes, but are content to estimate the magnitude of 
corrections one may expect. So we stay with $N_s^3$ volumes and focus 
on results in the continuum limit for 
\begin{equation} \label{Ls}
  L_s = a N_s = (5-10)\ {\rm fermi}\ .
\end{equation}
Finite volume corrections to the infinite volume continuum limit are
expected to be relevant as long as the volume is not large compared 
to a typical hadronic correlation length, which is about one fermi. 
For relatively small volumes an appropriate modeling 
of the BCs is necessary.

In the next section we introduce BCs, which reflect a (very) low 
temperature outside of the deconfined region. Two constructions, the 
``disorder wall'' and the ``confinement wall'' are carried out and 
shown to exhibit the usual asymptotic scaling properties in the finite 
volume continuum limit. While the confined wall is physically more 
accurate, the disorder wall can be easily implemented in MC 
calculations. For the latter MC calculations of pure SU(3) LGT
are performed in section~\ref{SU3_MC} and evidence for scaling 
is already found for systems sizes which can be simulated on PC 
clusters. Summary and conclusion are given in section~\ref{sec_Sum}.

\section{Boundary Conditions \label{sec_BC}}

Statistical properties of a quantum system with Hamiltonian $H$ in 
a continuum volume $V$, which is in equilibrium with a heatbath at 
physical temperature $T$, are determined by the partition function
\cite{Rothe}
\begin{equation}\label{e:Z}
    Z(T,V)={\rm Tr} e^{-H/T}=
    \sum_{\phi}\langle\phi|e^{-H/T}|\phi\rangle,
\end{equation}
where the sum extends over all possible states $|\phi\rangle$ of the 
system and the Boltzmann constant is set to one. Imposing periodic 
boundary conditions in Euclidean time $\tau$ and bounds of integration 
from $0$ to $1/T$, one can rewrite the partition function (\ref{e:Z})
in the path integral representation:
\begin{equation}\label{e:Zpath}
    Z(T,V)=\int D\phi\exp\left\{-\int_0^{1/T}d\tau L_E
    (\phi,\dot\phi)\right\}.
\end{equation}
Nothing in this formulation requires to carry out the infinite volume 
limit. In the contrary, if one deals with rather small volumes for 
which fluctuations of the mean values are not negligible, the idea
to consider $V\to\infty$ instead of $V$ appears to be rather obscure.

An obvious problem of applying equilibrium thermodynamics to deconfined 
volumes at RHIC is that there is no heatbath in sight with which the 
system could be in equilibrium. However, arguments have been made in 
the literature that after the rapid heating quench, when the deconfined 
volume is at about its maximum size, a (pseudo) equilibrium state is 
reached for a transitional period, which is reasonably long on the 
scale of the relaxation times involved. These arguments are not beyond 
doubts (some are raised in one of our own papers \cite{Ba06}), but 
they are not a topic of our present work. Here our assumption is that 
there is some truth to the belief that the bulk properties can be 
described by equilibrated finite temperature QCD. 

In MC simulation the updating process provides the heatbath and 
finite volumes can be equilibrated with all kind of BCs imposed,
although in practice most simulations have used periodic BCs. For 
simplicity and to be definite we restrict our discussion to pure 
SU(3) LGT. Generalization of the arguments of this section to full 
QCD appears to be straightforward. We use the Wilson action given by
\begin{equation} \label{action}
  S(\{U\}) = \frac{\beta^g}{3} \sum_{\Box} {\rm Re}\,
  {\rm Tr} \left( U_{\Box}\right)\,,
\end{equation}
$ U_{\Box} = U_{i_1j_1} U_{j_1i_2} U_{i_2j_2} U_{j_2i_1}$, where the 
sum is over all plaquettes of a 4D simple hypercubic lattice, $i_1,\,
j_1,\,i_2$ and $j_2$ label the sites circulating about the plaquette 
and $U_{ji}$ is the $SU(3)$ matrix associated with the link $\langle 
ij\rangle$. The reversed link is associated with the inverse matrix, 
and $\beta^g$ is related to the bare coupling constant by $\beta^g=6/
g^2$. The theory is defined by the expectation values of its operators 
with respect to the Euclidean path integral 
\begin{equation} \label{Z}
  Z = \int \prod_{\langle ij\rangle} dU_{ij}\,
      e^{S\left(\{U\}\right)}\,,
\end{equation}
where the integrations are over the invariant group measure, which was
for compact groups like SU(3) introduced by Hurwitz \cite{Hu} (Haar 
\cite{Ha} generalized it later).

Numerical evidence suggests that for $N_{\tau}$ fixed and $N_s\to\infty$ 
SU(3) lattice gauge theory exhibits a deconfining phase transition 
(sub- or superscript $t$ stand for transition) at some coupling 
$\beta^g_t(N_{\tau})=6/g^2_t (N_{\tau})$, which is weakly first 
order~\cite{firstorder}. The scaling behavior of the deconfining 
temperature is 
\begin{equation} \label{Tc_scaling}
  T^c = c_T\,\Lambda_L
\end{equation}
where $c_T$ is a constant and we use the lambda lattice scale 
\begin{equation} \label{Lambda}
  a\,\Lambda_L = f_{\lambda}(\beta^g) = \lambda(g^2)\,
  \left(b_0\,g^2\right)^{-b_1/(2b_0^2)}\,e^{-1/(2b_0\,g^2)}\,,
\end{equation}
where $a$ is the lattice spacing. The coefficients $b_0$ and $b_1$ are 
perturbatively determined by the renormalization group equation and 
independent of the renormalization prescription \cite{AF}, 
\begin{equation} \label{b_cof}
  b_0 = \frac{11}{3}\frac{3}{16\pi^2}~~{\rm and}~~
  b_1=\frac{34}{3}\left(\frac{3}{16\pi^2}\right)^2\,. 
\end{equation}
For perturbative and non-perturbative corrections we adopt the analysis 
of \cite{Bo96} in the parameterization of \cite{Ba06a}:
\begin{equation} \label{lambda}
  \lambda(g^2)\ =\ 1+a_1\,e^{-a_2/g^2}+a_3\,g^2+a_4\,g^4 
\end{equation}
with $a_1=71553750$, $a_2=19.48099$, $a_3=-0.03772473$, and
$a_4=0.5089052$.

We want to model an equilibrium situation surrounded by cold boundaries 
at $300\,$K. The effects are expected to penetrate at least a few 
correlation lengths into the volume. Note that this is the correlation 
length set by hadronic interaction, which can be defined by an 
appropriate inverse mass. On the lattice $\xi/a$ governs the continuum 
limit. This should not be confused with correlations due to the phase 
transition. As the lattice regularization allows to construct finite 
volume continuum limits as well as the infinite volume continuum limit, 
the question is: Which construction has the best chances to capture 
the physics realistically for a quark-gluon plasma in a small volume 
as typically created at RHIC? 

Let us first exemplify that LGT thermodynamics allows not only to 
approach the infinite volume continuum limit, but also finite 
volume continuum limits. For instance, we can define the thermodynamics 
on a torus which has the volume of, say, $(10\,{\rm fermi})^3$. To 
achieve a continuum limit, we have to send the lattice spacing $a\to 0$ 
in units of the physical scale. This is governed by the renormalization 
group equation and requires infinitely many lattice points $N_s = L_s/a
\to \infty$, while the physical volume of the lattice stays finite by 
arranging $L_s/{\rm fermi}=c_1$, where $c_1$ is a constant, e.g., 
$c_1=10$. The temperature of such a system is regulated by choosing 
$L_{\tau}/L_s=c_2$, where $c_2$ is a second constant. For describing 
deconfined volumes at RHIC the thus defined toroidal mini-universe 
is even less suitable than the conventionally used infinite volume 
limit, because correlations are artificially propagated through the 
periodic BCs. We present now two constructions of BCs, which reflect
cold exterior volumes. The second is physically more realistic, but
numerically more difficult to implement.

\subsection{Disorder wall \label{sub_Dwall}}

Imagine an almost infinite space volume $V=L_s^3$, which may have 
periodic BCs, and a smaller (very large, but small compared to $V$) 
sub-volume $V_0=L_{s,0}^3$. The complement to $V_0$ in $V$ will be 
called $V_1$. The number of temporal lattice links $N_{\tau}$ is the
same for both volumes. We want to find $\beta^g$ values and lattice 
dimensions so that scaling holds, while $V_0$ is at a temperature 
of $T_0=174\,{\rm MeV}$ and $V_1$ at $T_1=300\,{\rm K}$. We set the 
coupling $\beta^g$ to $\beta^g_0$ for plaquettes in $V_0$ and to 
$\beta^g_1$ for plaquettes in $V_1$. For that purpose any plaquette 
touching a site in $V_1$ is considered to be in $V_1$. Let us take 
$\beta^g_1=5.7$, which is at the beginning of the SU(3) scaling 
region. We have $174\,{\rm MeV} = 2.02 \times 10^{12}\, {\rm K}$ 
and, therefore,
\begin{equation}\label{e:Tratio}
  \frac{T_0}{T_1} = \frac{2.02\times 10^{12}}{300} = \frac{a_1}{a_0} 
  = \frac{f_{\lambda}(\beta^g_1)}{f_{\lambda}(\beta^g_0)}
\end{equation}
where $a_i$ is the lattice spacing in $V_i,\, i=0,1$.  Using 
(\ref{Lambda}) the scaling equation 
\begin{equation}\label{e:scaling}
   300\,f_{\lambda}(\beta^g_1) =
   2.02\times 10^{12}\,f_{\lambda}(\beta^g_0)
\end{equation}
yields $\beta^g_0=24.496$. $T_c$ estimates from MC calculations of
the literature extrapolate then to $L_{\tau}\ge 2.74\times 10^{10}\,a$ 
for the temporal lattice extension needed for a deconfined phase at 
$\beta^g_0=24.5$. This illustrates the orders of magnitude involved.

So, in practice we can only have $\beta^g_0$, but not $\beta^g_1$ in 
the scaling region, say $\beta^g_0=6$. There is still a reasonable 
choice for $\beta^g_1$. The scaling argument (\ref{e:Tratio}) shows 
that in lattice units a correlation length $\xi$, say one fermi, is 
at $\beta^g_0$ much larger than at $\beta^g_1$: 
\begin{equation}\label{e:xirat}
  \frac{(\xi/a_0)}{(\xi/a_1)} = \frac{a_1}{a_0} \approx 10^{10}\,.
\end{equation}
Now $\xi/a_0$ is of order one at $\beta^g_0=6$, so that $\xi/a_1$
becomes very small and the lowest order of the strong coupling 
expansion applies (see \cite{Rothe,Mu1} for the string tension and 
\cite{Mu2} for the glueball mass) in which $\xi/a_1 \sim -1/
\ln(\beta^g_1/18)$ so that $\beta^g_1\approx 10^{-10^{10}}$. We 
call this construction the disorder wall.

The disorder wall allows for a finite volume continuum limit, which 
is approached for $\beta^g_0\to\infty$, and the usual scaling law
holds. Eventually (obviously for $\beta^g_0>24$) the value of 
$\beta^g_1$ increases also, unless the outside temperature is 
exactly zero and, hence, $\beta^g_1=0$. An outside temperature 
of 300$\,$K is on the scale of the inside temperature so close 
to zero that in practical MC simulations the effects of the 
$\beta^g_1$ increase are not noticeable and $\beta^g_1=0$ is a 
safe choice independently of $\beta^g_0$. Simulation results in
section~\ref{SU3_MC} support that scaling in $\beta^g_0$ holds 
already for $\beta^g_0$ values, which can be reached in practice.

\subsection{Confinement wall \label{sub_Cwall}}

In the continuum limit the disorder wall separates regions in which 
a correlation length $\xi$ takes on entirely different magnitudes when 
measured in unit of the lattice spacing (\ref{e:xirat}). Although we 
are only interested in the physics inside the wall, a construction 
for which the distance of one fermi stays in lattice units continuous 
across the boundary is clearly more physical. Along similar lines as 
before this can be achieved by using an anisotropic lattice outside of
the deconfined region. Let us denote in $V_1$ the spacelike links by 
$a_s$, the timelike links by $a_{\tau}$, and use there the Wilson action
\begin{equation} \label{action_asym}
  S(\{U\}) = \frac{\beta^g_s}{3} \sum_{{\Box}_s} {\rm Re}\,
             {\rm Tr} \left( U_{{\Box}_s}\right) 
           + \frac{\beta^g_{\tau}}{3} \sum_{{\Box}_{\tau}} {\rm Re}\, 
             {\rm Tr} \left( U_{{\Box}_{\tau}}\right)\,,
\end{equation}
where $\beta^g_s$ and $\beta^g_{\tau}$ are the couplings of the 
spacelike and timelike plaquettes, respectively. The lambda scale 
of this action has been investigated by Karsch \cite{Ka82} and in 
the continuum one finds~\cite{Karsch}
\begin{equation} \label{beta_ratio}
  \frac{\beta^g_{\tau}}{\beta^g_s} =
  \left(\frac{a_s}{a_{\tau}}\right)^2\,.
\end{equation}
As we aim at 
\begin{equation} \label{as}
  a_0 = a_s \approx 10^{-10}\,a_{\tau}
\end{equation}
the resulting orders of magnitude are even more astronomical than 
before. The sublattice $V_1$ is again driven out of the scaling 
region, which would be reached for sufficiently large values of 
$\beta^g_{\rm Karsch}=\sqrt{\beta^g_s\beta^g_{\tau}}$. For all 
practical purposes we are driven into the strong coupling region
and can set $\beta^g_{\tau}=0$, so that the 
simulation of the confined world becomes effectively 3D. By 
measuring an appropriate correlation length, for instance via the 
string tension or glueball mass, we can non-perturbatively tune 
$\beta^g_s$, so that $a_s=a_0$ holds. A MC simulation has then 
to include an outside world with a non-zero $\beta^g_s$. This is 
physically quite interesting, but computationally more demanding 
than using the disorder wall, for which we present MC simulations 
in the following.

\section{SU(3) MC Simulations with the Disorder Wall\label{SU3_MC}}

In the Monte Carlo calculations of this section we approximate a cold 
exterior by using the disorder wall BCs. In practice this means, we 
simply omit plaquettes, which involve links through the boundary. So 
we drop the subscript in the $\beta^g_0$ definition of the previous
section and return to simply using the $\beta^g$ notation. For both 
periodic and disorder wall BCs we present an analysis of data from 
simulations on $N_{\tau}\times N_s^3$ lattices with $N_{\tau}=4$ 
and~6.  The $N_s$ values and our statistics in sweeps are compiled 
in tables~\ref{tab_4stat} and~\ref{tab_6stat}.

As in \cite{Bo96} we use the maxima of the Polyakov loop susceptibility
\begin{equation}\label{chi_def}
  \chi_{\max}=\frac{1}{N_s^3}\left[\langle|P|^2\rangle-
  \langle|P|\rangle^2\right]_{\max},\,\,\,P=\sum_{\vec{x}}P_{\vec{x}}
\end{equation}
to define pseudo-transition couplings $\beta^g_{pt}(N_s;N_{\tau})$. 
For periodic BCs, indicated by the superscript $p$ of $a^p_3$, they 
have a finite size behavior of the form
\begin{equation} \label{beta_pbc}
  \beta^g_{pt}(N_s;N_{\tau}) = \beta^g_t(N_{\tau}) + 
  a_3^p\,\left(\frac{N_{\tau}}{N_s}\right)^3 +\ \dots\ .
\end{equation}
Fits to this form yield $\beta^g_t(N_{\tau})$ and estimates are given 
in Boyd et al.~\cite{Bo96}. Our estimates for $\beta^g_{pt}(N_s;4)$ are
summarized in table~\ref{tab_4stat}. Within statistical errors the 
lattice size dependence of $\beta^g_{pt}(N_s;4)$ is almost negligible.
A fit of our data with periodic BCs to (\ref{beta_pbc}) gives 
$\beta^g_t(4)=5.69236\,(21)$ in agreement with the value $5.6925\,(2)$ 
reported in \cite{Bo96}. From now on we use $\beta^g_t(4)=5.69236\,(21)$ 
as the infinite volume limit for periodic and disorder wall BCs.

\begin{table}
\centering
\begin{tabular}{|c|c|c|c|c|}
\hline 
 \multicolumn{1}{|c|}{ }     & \multicolumn{2}{c|}{Periodic BCs}  &
        \multicolumn{2}{c|}{Disorder wall BCs}   \\ \hline
$N_s$&$N_{meas}$    &  $\beta^g_{pt}(N_s;4)$  & $N_{meas}$    
     &$\beta^g_{pt}(N_s;4)$ \\ \hline
 12  &$32\times 10$\,$000$&$5.6904\,(27)$& $64\times 20\,000$&
$6.110\,(34)$\\ \hline
 16  &$32\times 10$\,$000$&$5.6912\,(11)$&$64\times 20\,000$&
$5.8460\,(83)$ \\ \hline
 20  &$32\times 10$\,$000$&$5.69184\,(69)$&$32\times 10\,000$&
$5.7744\,(59)$\\ \hline
 24  &$32\times 10$\,$000$&$5.69170\,(41)$&$32\times 10\,000$&
$5.7426\,(30)$\\ \hline
 32  &$32\times 10$\,$000$&$5.69225\,(16)$&$32\times 10\,000$&
$5.7192\,(11)$\\ \hline
\end{tabular} 
\caption{Number of measurements and pseudo-transition coupling 
estimates for $4\times N_s^3$ lattices.} \label{tab_4stat}
\end{table} 

Our estimates of $\beta^g_{pt}(N_s;6)$ are summarized in 
table~\ref{tab_6stat}. We took only few $N_{\tau}=6$ data for 
periodic BCs, because they consume already considerable CPU
time and results exist already in the literature. Again they show 
almost no lattice size dependence. Fitting them to (\ref{beta_pbc}) 
gives $\beta^g_t(6)= 5.8926\,(18)$ in agreement with $\beta^g_t(6)=
5.8941\,(5)$ from \cite{Bo96}. In the following we use the latter,
more accurate, $\beta^g_t(6)$ estimate of the literature.

\begin{table}
\centering
\begin{tabular}{|c|c|c|c|c|}
\hline 
 \multicolumn{1}{|c|}{ }     & \multicolumn{2}{c|}{Periodic BCs}  &
        \multicolumn{2}{c|}{Disorder wall BCs}   \\ \hline
$N_s$&$N_{meas}$    &  $\beta^g_{pt}(N_s;6)$  & $N_{meas}$    
     &$\beta^g_{pt}(N_s;6)$ \\ \hline
 18  &$32\times 10$\,$000$&$5.8932\,(48)$& $32\times 10\,000$&
$6.47\ (14)$\\ \hline
 20  &$-$&$-$& $32\times 10\,000$& $6.27\ (04)$\\ \hline
 24  &$32\times 10$\,$000$&$5.8934\,(26)$& $32\times 10\,000$&
$6.089\,(23)$\\ \hline
 28  &$-$&$-$& $32\times 10\,000$& $6.012\,(11)$\\ \hline
 32  &$32\times 10$\,$000$&$5.8927\,(12)$& $32\times 10\,000$&
$5.9812\,(73)$\\ \hline
 40  &$-$&$-$& $32\times 10\,000$& $5.9463\,(53)$\\ \hline
 48  &$-$&$-$& $16\times 12\,000$& $5.9271\,(38)$\\ \hline
\end{tabular} 
\caption{Number of measurements and pseudo-transition coupling 
estimates for $6\times N_s^3$ lattices.} \label{tab_6stat}
\end{table} 

The disorder wall BCs introduce an order $N_s^2$ disturbance, so that 
Eq.~(\ref{beta_pbc}) becomes
\begin{eqnarray} \label{beta_cbc}
  \beta^g_{pt}(N_s;N_{\tau}) &=& \beta^g_t(N_{\tau}) + 
  a_1^d\,\frac{N_{\tau}}{N_s}\\ \nonumber &+& 
  a_2^d\,\left(\frac{N_{\tau}}{N_s}\right)^2 +
  a_3^d\,\left(\frac{N_{\tau}}{N_s}\right)^3 +\ \dots\ ,
\end{eqnarray}
where the superscripts $d$ of the coefficients $a_i^d$ indicate
disorder wall BCs.

\begin{figure}[-t] \begin{center} 
\epsfig{figure=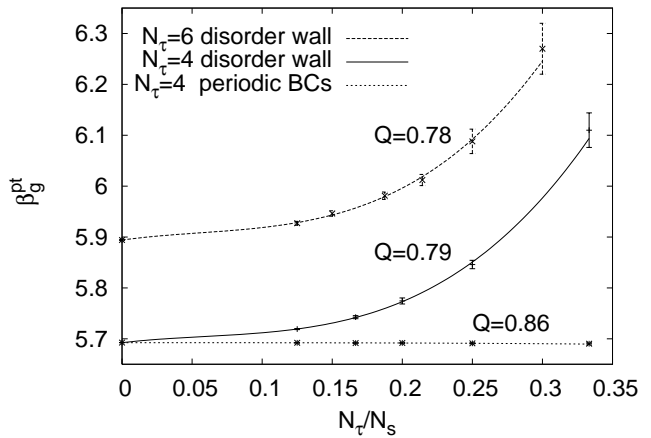,width=\columnwidth} \vspace{-1mm}
\caption{Fits of pseudo-transition coupling constant values and
their infinite volume extrapolations.  \label{fig_TcFits} }
\end{center} \vspace{-3mm} \end{figure}

In Fig.~\ref{fig_TcFits} we show the fit (\ref{beta_cbc}) to 
$\beta^g_t(4)$ and our pseudo-transition values $\beta^g_{pt}(N_s;4)$ 
from simulations with the disorder wall BCs. The very precise infinite 
volume estimate $\beta^g_t(4)$ from simulations with periodic BCs
is included in the disorder wall data to stabilize the fit at large 
volumes. For comparison the fit (\ref{beta_pbc}) for our $N_{\tau}=4$ 
data from simulations with periodic BCs is also given. While finite 
size corrections are practically negligible for the simulations with 
periodic BCs, this is not the case for the disorder wall BCs. $Q$ 
is the goodness of the fit (e.g., chapter~2.8 of \cite{BBook}).

We also include the fit (\ref{beta_cbc}) to $\beta^g_t(6)$ and 
our $\beta^g_{pt}(N_s;6)$ disorder wall data in Fig.~\ref{fig_TcFits}. 
As expected both disorder wall curves show strong finite lattice size 
effects. This is not automatically of physical relevance. Important is
whether universal corrections survive in the finite volume continuum 
limit. Here universal means that the corrections do not depend on the 
lattices used in the simulations, once these lattices are sufficiently
large. In the following we test this for $N_{\tau}=4$ and $N_{\tau}=6$ 
using the lambda scale (\ref{Lambda}) to calculate estimates in 
physical units. Although our $\beta^g$ values used are rather small, 
they are in the previously reported \cite{Bo96} scaling region for 
(\ref{Lambda}), so that it is reasonable to expect universal behavior 
with moderate corrections.

\begin{figure}[-t] \begin{center} 
\epsfig{figure=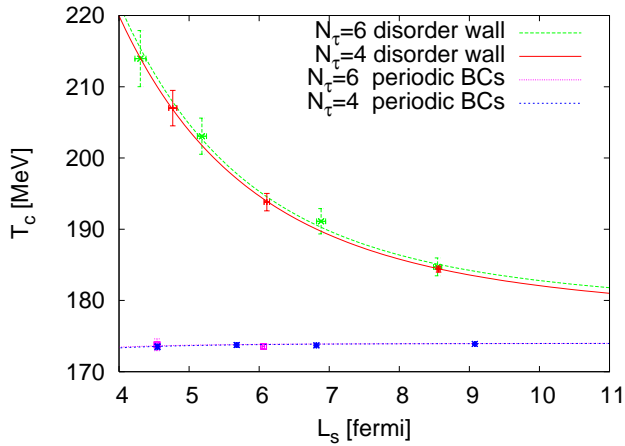,width=\columnwidth} \vspace{-1mm}
\caption{Estimate of finite volume corrections to the deconfinement
temperature, set at 174$\,$MeV for an infinite volume.
\label{fig_boxTc} } 
\end{center} \vspace{-3mm} \end{figure}

The infinite volume $T^c$ value (\ref{Tc}), the lambda scale 
(\ref{Lambda}) and Eq.~(\ref{Tc_scaling}) give us the $g^2$ dependence 
of the lattice spacing $a$ in units of fermi. Using the $N_{\tau}=4$ 
fit to (\ref{beta_cbc}) of Fig.~\ref{fig_TcFits} together with 
Eqs.~(\ref{LGT_T}) and (\ref{Ls}) allows to eliminate $g^2$ and we 
plot the resulting function $T^c(L_s)$ in Fig.~\ref{fig_boxTc}. 
Repeating this procedure for our $N_{\tau}=6$ fit to (\ref{beta_cbc}) 
gives, as is seen in Fig.~\ref{fig_boxTc}, almost the same $T^c(L_s)$ 
dependence and thus provides some evidence that our $N_{\tau}=4$ 
and~6 results are already representative for the finite volume 
continuum limit. For a box of volume $(10\,{\rm fermi})^3$ 
the pseudocritical temperature $T^c$ is about 5\% higher than the 
infinite volume estimate and this correction increases to about 17\% 
for a $(5\,{\rm fermi})^3$ box. For comparison we use the same 
procedure for analyzing the finite size dependence obtained by 
fitting the $N_{\tau}=4$ and~6 data with periodic BCs (including and 
enforcing the $\beta^g_t(6)=5.8941$ limit for $N_{\tau}=6$) to 
Eq.~(\ref{beta_pbc}).  This gives the two lower curves of the figure, 
which fall almost on top of one another. In that case error bars are 
considerably smaller than for the disorder wall data.

\begin{table}
\centering
\begin{tabular}{|c|c|c|c|c|} 
\hline
 \multicolumn{1}{|c|}{ }     & \multicolumn{2}{c|}{Periodic BCs}  &
        \multicolumn{2}{c|}{Disorder wall BCs}   \\ \hline
$N_s$ &$\chi_{\max}$&$\Delta\beta^g_{2/3}$  
      &$\chi_{\max}$&$\Delta\beta^g_{2/3}$ \\ \hline
 12   &3.585 (71)  &0.0207 (11) &1.715 (27)& 0.448 (18)  \\ \hline
 16   &7.62 (16)   &0.0103 (13) &1.879 (26)& 0.0997 (21) \\ \hline
 20   &16.17 (67)  &0.00498 (68)&2.525 (84)& 0.0440 (27) \\ \hline
 24   &28.6 (1.1)  &0.00277 (32)&3.58 (14) & 0.0225 (16) \\ \hline
 32   &73.0 (2.0)  &0.00132 (16)&7.67 (39) & 0.00709 (61)\\
\hline
\end{tabular} 
\caption{Maxima of Polyakov loop susceptibility and width of
the transition for lattices $4\times N_s^3$.}\label{tab_4dbeta}
\end{table} 

\begin{figure}[-t] \begin{center}
\epsfig{figure=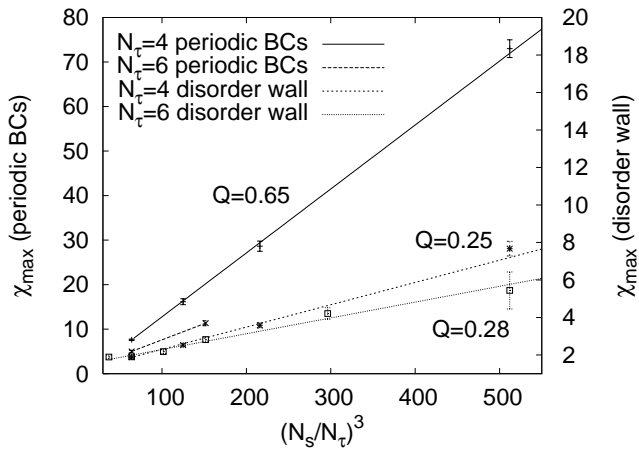,width=\columnwidth} \vspace{-1mm}
\caption{Maxima of Polyakov loop susceptibility.}\label{fig_chi_max}
\end{center} \vspace{-3mm} \end{figure}

For a first order phase transition the maxima of the Polyakov loop
susceptibility have to scale with the system volume to reproduce 
the delta function like singularity of a first order transition. 
For $N_{\tau}=4$ our $\chi_{\max}$ values are listed in 
table~\ref{tab_4dbeta} and for $N_{\tau}=6$ in table~\ref{tab_6dbeta}. 
Using $N_s\ge 16$ for $N_{\tau}=4$ and $N_s\ge 20$ for $N_{\tau}=6$ 
acceptable fits to the straight-line form 
\begin{equation} \label{fit_chimax}
  \chi_{\max} =d_1+d_2\,(N_s/N_{\tau})^3 
\end{equation}
are obtained and shown together with their $Q$ values in 
Fig.~\ref{fig_chi_max} (for $N_{\tau}=6$ and periodic BCs only
two data points are fitted, so that there is no $Q$ value in that 
case). To enhance the scale for the disorder wall fits, they are 
displayed on the right ordinate. The leading coefficients obtained 
from fits for disorder wall data differ from those for the data 
with periodic BCs: $d_2^d=0.01190\,(60)$ versus $d_2^p= 0.1436\,(37)$ 
for $N_{\tau}=4$ and $d_2^d=0.00843\,(95)$ versus $d_2^p= 0.0723\,(70)$ 
for $N_{\tau}=6$. This is possible because the Polyakov loop maxima are 
not physical observables, but bare quantities.

\begin{table}
\centering
\begin{tabular}{|c|c|c|c|c|} 
\hline
 \multicolumn{1}{|c|}{ }     & \multicolumn{2}{c|}{Periodic BCs}  &
        \multicolumn{2}{c|}{Disorder wall BCs}   \\ \hline
$N_s$ &$\chi_{\max}$&$\Delta\beta^g_{2/3}$  
      &$\chi_{\max}$&$\Delta\beta^g_{2/3}$ \\ \hline
 18   &  2.47 (10) &0.0303 (21) &2.15 (20)& 0.84  (04)  \\ \hline
 20   &  $ - $     &$-$         &1.89 (10)& 0.322 (25)  \\ \hline
 24   &  5.00 (27) &0.0162 (15) &1.90 (11)& 0.140 (13)  \\ \hline
 28   &  $ - $     &$-$         &2.18 (10)& 0.0712 (69) \\ \hline
 32   & 11.34 (55) &0.00803 (66)&2.82 (16)& 0.0422 (37) \\ \hline
 40   &  $ - $     &$-$         &4.21 (31)& 0.0234 (17) \\ \hline
 48   &  $ - $     &$-$         &5.44 (99)& 0.0123 (38) \\ \hline
\hline
\end{tabular} 
\caption{Maxima of Polyakov loop susceptibility and width of
the transition for lattices $6\times N_s^3$.}\label{tab_6dbeta}
\end{table} 

\begin{figure}[-t] \begin{center}
\epsfig{figure=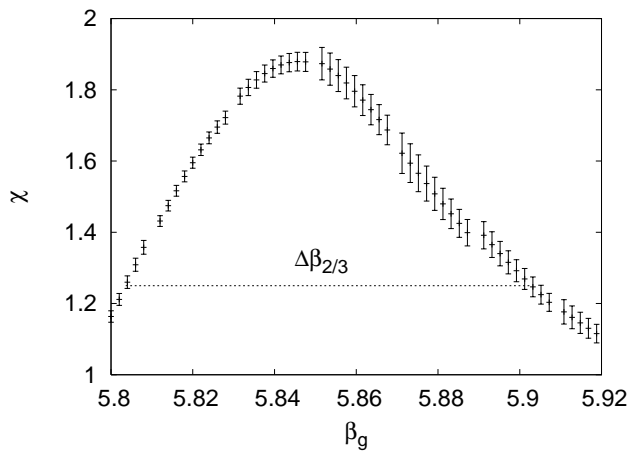,width=\columnwidth} \vspace{-1mm}
\caption{The Polyakov loop susceptibility on a $4\times16^3$ lattice with
disorder wall BCs.  \label{fig_chi_4x16} }
\end{center} \vspace{-3mm} \end{figure}

Let us now consider the width of the transition. For a $4\times16^3$ 
lattice with disorder wall BCs the Polyakov loop susceptibility as a 
function of $\beta^g$ is shown in Fig.~\ref{fig_chi_4x16}. We use 
reweighting \cite{FeSw88,BBook} to cover a range of $\beta^g$ and define 
$\Delta\beta^g_{2/3}$ as the width of a peak at $2/3$ of its height. 
On large lattices the figures look less nice than Fig.~\ref{fig_chi_4x16},
but provide still sufficient information to extract $\Delta\beta^g_{2/3}$.
Our estimates for $N_{\tau}=4$ are listed in table~\ref{tab_4dbeta} and 
for $N_{\tau}=6$ in table~\ref{tab_6dbeta}. The data are fitted to the form
\begin{equation}\label{dbeta_pbcfit}
  \Delta\beta^g_{2/3}=c_1^p\left(\frac{N_\tau}{N_s}\right)^3
                   +c_2^p\left(\frac{N_\tau}{N_s}\right)^6
\end{equation}
for periodic BCs and to
\begin{equation}\label{dbeta_cbcfit}
  \Delta\beta^g_{2/3}=c_1^d\left(\frac{N_\tau}{N_s}\right)^3
                   +c_2^d\left(\frac{N_\tau}{N_s}\right)^4
\end{equation}
for the disorder wall BCs. The first term reflects in both cases the 
delta function singularity of a first order phase transition. The 
leading order corrections to that differ due to the influence of the
BCs. 

\begin{figure}[-t] \begin{center}
\epsfig{figure=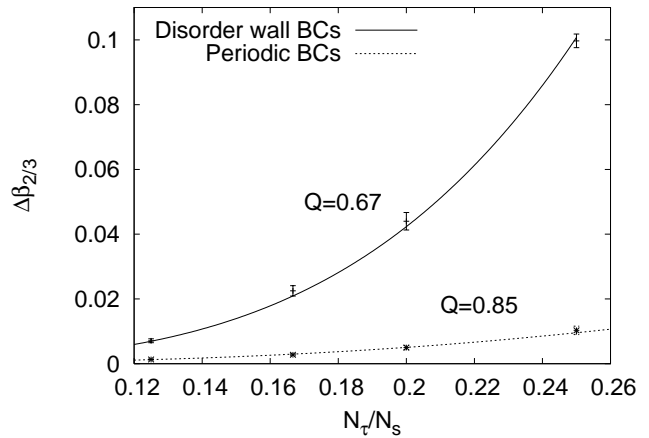,width=\columnwidth} \vspace{-1mm}
\caption{Fits of the $N_{\tau}=4$ width of the transition. 
\label{fig_4dbeta} } 
\end{center} \vspace{-3mm} \end{figure}

For $N_{\tau}=4$ the final fits (see below) are shown in 
Fig.~\ref{fig_4dbeta}. From the disorder wall data we have 
omitted our smallest $4\times12^3$ lattice from the fit, because 
the width becomes for it so broad that it spoils $Q$ (larger lattices 
than with periodic BCs are needed). The leading order coefficients 
are then $c_1^p=1.17\, (55)$ and $c_1^d=0.650\,(49)$. Both data sets 
together can still be consistently fitted using the weighted average 
$c_1=0.654\,(49)$ of the leading order coefficients and 1-parameter 
fits for $c_2$ in (\ref{dbeta_pbcfit}) and (\ref{dbeta_cbcfit}). This 
ensures that the ratio of the widths becomes one in the infinite 
volume limit. In contrast to the Polyakov loop maxima, the width 
of the transition is a physical observable, which is to leading 
order in the volume independent of the BCs. We have chosen to 
show these 1-parameter fits together with their $Q$ values in 
Fig.~\ref{fig_4dbeta}. 

\begin{figure}[-t] \begin{center}
\epsfig{figure=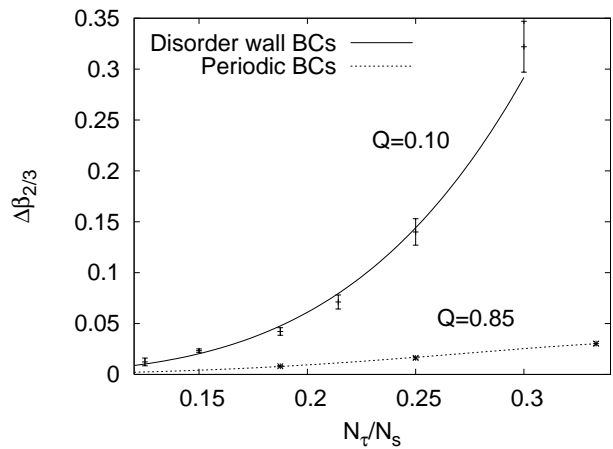,width=\columnwidth} \vspace{-1mm}
\caption{Fits of the $N_{\tau}=6$ width of the transition. 
\label{fig_6dbeta} } 
\end{center} \vspace{-3mm} \end{figure}

The methodology for the corresponding $N_{\tau}=6$ fits is the same 
as before. Using the data of table~\ref{tab_6dbeta} we find the leading 
order coefficients $c_1^p=1.27\, (11)$ and $c_1^d=1.8\,(1.3)$, which 
average to $c_1=1.27\, (11)$. With this values consistent 1-parameter 
fits are obtained and together with their $Q$ values depicted in 
Fig.~\ref{fig_6dbeta}. Note that the ordinate scale in this figure 
is more than three times larger than in Fig.~\ref{fig_4dbeta}. 
Nevertheless the extracted physical values have to be the same 
to the extent that scaling holds.

\begin{figure}[-t] \begin{center}
\epsfig{figure=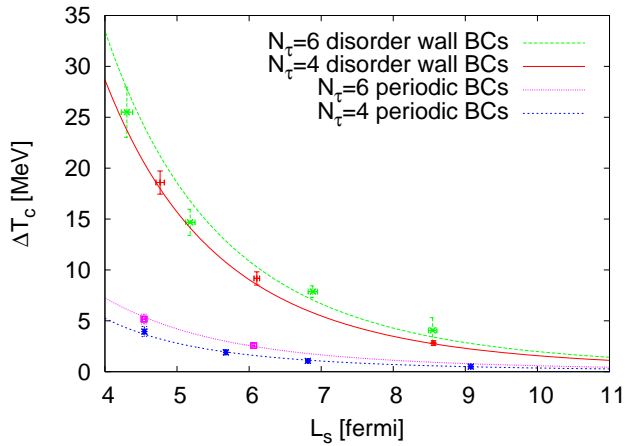,width=\columnwidth} \vspace{-1mm}
\caption{Estimate of finite volume correction to the width of the 
deconfinement phase transition. \label{fig_dTc} }
\end{center} \vspace{-3mm} \end{figure}

We want to plot the width in physical units of MeV versus the box size 
in fermi and follow a similar procedure as before for $T^c(L_s)$. For 
given $L_s(\beta^g)$ we define
\begin{equation}\label{dT}
  \Delta T^c(L_s) = T(\beta^g_{pt}+\Delta\beta^g_{2/3}/2) 
  - T(\beta^g_{pt}-\Delta\beta^g_{2/3}/2)\,.
\end{equation}
The dependence $\Delta T(L_s)$ is shown in Fig.~\ref{fig_dTc}. Compared 
to periodic BCs disorder wall BCs lead to a substantial broadening of 
the transition for the volumes considered: At $(10\,{\rm fermi})^3$ 
by a factor of 4.3 and at $(5\,{\rm fermi})^3$ by a factor of 5.5. 
The width is slightly less than 1\% of the (enhanced) transition 
temperature at $(10\,{\rm fermi})^3$ and about 8\% at 
$(5\,{\rm fermi})^3$. Within the error bars, which are quite
large for the widths, scaling works again well.

Simulations with disorder wall BCs have turned out to be far more CPU 
time consuming than those with periodic BCs. The decreased heights 
of our pseudo-transition signals (the maxima of the Polyakov loop 
susceptibilities), their increased widths and the strong finite 
size effects are the underlying reasons. While the reweighting range 
\cite{FeSw88,BBook} is about the same for simulations with periodic 
or disorder wall BCs on identically sized lattices, accurate disorder 
wall results require far more patches, i.e., independent simulations 
at distinct $\beta^g$ values. In addition the signal is worse due to 
the decreased heights of the peaks. Finally extrapolations to 
infinite lattices, as needed for the finite volume continuum limit, 
are more demanding due to the strong and more sophisticated finite 
size corrections. This is helped by including infinite volume 
extrapolations from periodic lattices as disorder wall data points, 
what can be done because these extrapolations do not depend on the 
BCs. Still, data from our largest $6\times 48^3$ lattice turn out 
to be essential to stabilize the $N_{\tau}=6$ fits for large lattice 
sizes. Far smaller lattices are sufficient when periodic BCs are used.

\section{Summary and Conclusions \label{sec_Sum}}

Relatively small physical volumes as typical for the deconfined 
phase in BNL RHIC \cite{RHIC} lead in our SU(3) investigation to 
a substantial rounding of the transition and to an increase of the 
(effective) deconfinement temperature by 5\% to 20\%. The physical 
reason for these corrections is that correlation lengths are 
proportional to $L_{\tau}$ and $L_{\tau}/L_s$ does not approach 
zero in the finite volume continuum limit. We estimate that the 
corrections are negligible for the geometry of a torus, used 
in practically all previous LGT simulations of the subject, but 
not for BCs which reflect the cold exterior. Using the disorder wall 
BCs introduced in section~\ref{sub_Dwall}, we find the magnitude of 
the effects on pure SU(3) LGT competitive to those of other 
corrections, foremost the inclusion of quarks. In particular the 
width of the transition increases by factors 4 to 6 over the width 
found for a torus of the same physical size.

Most scattering events at a RHIC are not from central collisions. 
So one has to cope with a distribution of volumes, each of it 
associated with its own effective deconfinement temperature and 
width. Due to such finite volume effects the concept of a sharp 
transition becomes blurred even when the effects of quarks, which 
convert the transition into a crossover \cite{Ao06}, are not yet 
taken into account.
Ultimately one may want to extend LGT studies of the deconfining
transition with BCs reflecting the confined outside world to full 
QCD. Before doing so, additional experience can be gained from pure 
SU(3) LGT. In future work we intend to perform simulation for the 
physically more realistic confinement wall BCs, which we introduced 
in section~\ref{sub_Cwall}. Further, the entire equilibrium 
thermodynamics of the deconfined phase ought to be addressed. 

Close to the transition point the deconfined phase can only be studied 
non-perturbatively. Equilibrium QCD on the lattice allows this from 
first principles. But the assumption that equilibrium configurations 
can capture the essence of the transition is a strong one. Studies of 
the important dynamical aspects of the transition have to rely on 
phenomenological approaches. For instance, models which use 
hydrodynamics in the quark gluon plasma stage reproduce experimental 
data on particle abundances and flows well~\cite{pheno}.

\acknowledgments
This work was in part supported by the US Department of Energy 
under contract DE-FG02-97ER41022.

\clearpage

\begin{thebibliography}{19}

\bibitem{Tc-old} L.D. McLerran and B. Svetitsky, Phys. Lett. B
{\bf 98}, 195 (1981); 
J. Kuti, J. Pol\'onyi, and K. Szlach\'anyi, Phys. Lett. B {\bf 98},
199 (1981). 

\bibitem{Tc-now} P. Petreczky, Nucl. Phys. B (Proc. Suppl.) {\bf 140},
78 (2005);  E. Laermann and O. Philipsen, Ann. Rev. Nucl. Part. Sci.
{\bf 53}, 163 (2003). 

\bibitem{quantum} We omit the adverb ``quantum'' in front of continuum
limit from hereon. The classical continuum limit is not considered in 
our paper.

\bibitem{Rothe} H.J. Rothe, {\it Lattice Gauge Theories: An 
Introduction}, 3rd edition, World Scientific, Singapore 2005.

\bibitem{note} Often the physical scale in pure gauge calculations
is set by choosing a value of about 420$\,$MeV for the string 
tension $\sqrt{\sigma}$, which then leads via scaling to $T^c\approx
265\,$MeV. However, in this paper we prefer to deal with a $T^c$
value close to its physical QCD estimate, as one is ultimately 
interested in corrections to this value.

\bibitem{RHIC} For a review see, e.g., B. Muller, J.L. Nagle, Ann. Rev. 
Nucl. and Part. Sci.  {\bf 56}, 93 (2006) 

\bibitem{Ba06} A. Bazavov, B. Berg, and A. Velytsky, Phys. Rev. D 
{\bf 74}, 014501 (2006).

\bibitem{Ba06a} Eq.~(19) of \cite{Ba06}.

\bibitem{Hu} A. Hurwitz, G\"ottinger Gesellschaft der Wissenschaften
{\bf 1897}, 71 (1897).

\bibitem{Ha} A. Haar, Ann. Math. {\bf 34}, 147 (1933).

\bibitem{firstorder} F. Brown, N. Christ, Y. Deng, M. Gao, and T. Woch,
Phys. Rev. Lett. {\bf 61}, 2058 (1988); M. Fukugita, M. Okawa, A.
Ukawa, Phys. Rev. Lett. {\bf 63}, 1768 (1989); N.A. Alves, B.A. Berg, 
and S. Sanielevici, Phys. Rev. Lett. {\bf 64}, 3107 (1990).

\bibitem{AF} D.J. Gross and F. Wilczek, Phys. Rev. Lett. {\bf 30}, 
1343 (1973); H.D. Politzer, Phys. Rev. Lett. {\bf 30}, 1346 (1973).

\bibitem{Bo96} G. Boyd, J. Engels, F. Karsch, E. Laermann, C. Legeland,
M. L\"utgemeier, and B. Petersson, Nucl. Phys. B {\bf 469}, 419 (1996).

\bibitem{Mu1} J. Kogut, R. Pearson, and J. Shigemitsu, Phys. Rev. Lett.
{\bf 43}, 484 (1979); G. M\"unster and P. Weisz, Phys. Lett. B {\bf 96},
119 (1980).

\bibitem{Mu2} G. M\"unster, Nucl. Phys. B {\bf 190} [FS3], 439 (1981);
Erratum {\bf 200} [FS4], 536 (1982).

\bibitem{Ka82} F. Karsch, Nucl. Phys. B {\bf 205} [FS5], 285 (1982).

\bibitem{Karsch} This follows from Eq.~(1.4) of \cite{Ka82}. Note that
in that equation $g^2_{\sigma}(a,\xi)/g^2_{\tau}(a,\xi)\to 1$ in the 
continuum limit due to Eq.~(2.4) of \cite{Ka82}.

\bibitem{BBook} B.A. Berg, {\it Markov Chain Monte Carlo Simulations 
and Their Statistical Analysis}, World Scientific, Singapore, 2004.

\bibitem{FeSw88} A.M. Ferrenberg and R.H. Swendsen, Phys. Rev. Lett.
{\bf 61}, 2635 (1988); erratum {\bf 63}, 1658 (1989).

\bibitem{Ao06} Y. Aoki, G. Endr\'odi, Z. Fodor, S.D. Katz, and 
K.K.  Szab\'o, Nature {\bf 443}, 675 (2006).

\bibitem{pheno} D. Teaney, J. Lauret, and E.V. Shuryak, Phys. Rev. 
Lett. {\bf 86}, 4783 (2001); T. Hirano and M. Gyulassy, Nucl. Phys.
A {\bf 769}, 71 (2006).

\end{thebibliography}
\end{document}